# Controlled co-excitation of direct and indirect ultrafast demagnetization in Co/Pd multilayer with large perpendicular magnetic anisotropy

Santanu Pan[1], Olav Hellwig[2], and Anjan Barman[1,*]

[1]*Department of Condensed Matter Physics and Material Sciences, S N Bose National Centre for Basic Sciences, Block JD, Sector III, Salt Lake, Kolkata 700106, India*

[2]*Institute of Physics, Chemnitz University of Technology, Reichenhainer Straße 70, D-09107 Chemnitz,Germany and Institute of Ion Beam Physics and Materials Research, Helmholtz-Zentrum Dresden-Rossendorf, 01328 Dresden, Germany*

(Dated: November 30, 2018)

Ever since its discovery in 1996, ultrafast demagnetization has ignited immense research interest due to its scientific rigor and technological potential. A flurry of recent theoretical and experimental investigations has proposed direct and indirect excitation processes in separate systems. However, it still lacks a unified mechanism and remains highly debatable. Here, for the first time, we demonstrate that instead of either direct or indirect interaction, simultaneous and controlled excitation of both direct and indirect mechanisms of demagnetization are possible in a multilayers composed of repeated Co/Pd bi-layers. Moreover, we were able to modulate demagnetization time (from ~350 fs to ~750 fs) by fluence and thickness dependent indirect excitation due to heat current flowing vertically downward from top layers, which is combined with an altogether different scenario of direct irradiation. Finally, by regulating the pump wavelength we could effectively control the contribution of indirect process, which gives a confirmation to our understanding of the ultrafast demagnetization process.

## I. INTRODUCTION

Since the discovery of ultrafast demagnetization, more than twenty years ago in the pioneering experiment by Beaurepaire *et al.*[1], it has become a hot topic in magnetism research[2-9]. However, technological application in spintronics applications demand a prior understanding of the underlying microscopic mechanism which is found to be intriguing as well as challenging. This challenge is far more intense for complex systems, such as alloys and multilayers. A wide range of theoretical[9-14] and experimental[15-24] investigations have been brought into the picture over the years to explain the underlying mechanism of this ultrafast modification of magnetization. Most of the results claim a direct interaction between a laser pulse and the ferromagnetic material, and are based on spin-flip scattering (SFS) resulting from spin-orbit interaction, such as Elliott-Yafet like electron-phonon scattering, electron-magnon scattering, Coulomb exchange scattering, and relativistic spin-flip scattering. In 2010 a microscopically different theoretical proposal by Battiato *et al*.[12] followed up by several experimental observations[16-18,20] demonstrated that laser-excited hot electrons play a crucial role in ultrafast demagnetization through spin-dependent transport or heat current transport. Although the role of spin current and diffusive heat flow is controversial, it could convincingly explain the process of ultrafast demagnetization without any consideration of SFS processes, where direct interaction is not the primary mechanism. So far, the experimental demonstrations of indirect excitations involved either complicated experimental arrangements or a tricky alteration of magnetization states in different layers[23,25-28]. Here, using a very simple approach we clearly observe the presence of indirectly excited ultrafast demagnetization in a multilayer system, without using any additional source of spin current.

Although the phenomena of SFS (direct) and diffusive heat current flow (indirect) are very different in terms of microscopic mechanisms, they act on a similar time scale, which raises two serious questions. First, can both of these mechanisms together (direct and indirect) be responsible for the demagnetization in a sample? Second, if so, which one of those is more dominant and under which conditions? Recently, Turgut *et al*.[29] showed the presence of both spin-flip scattering and super diffusive spin current during demagnetization, in which the sample has been specially designed by changing the intermediate spacer layers. But, what it still lacks is the simultaneous direct and indirect excitation of ultrafast demagnetization for a simple and single system during a demagnetization process. Here, we have experimentally observed the simultaneous presence of both mechanisms controlling the demagnetization process. More importantly, we could control the individual contributions by changing the excitation fluence and sample thickness and hence, showed a transition from more direct process to a more indirect one.

## II. EXPERIMENTAL DETAILS

The experimental investigations and results presented in this article are performed on samples with the layer structure Ta(1.5)|Pd(3.0)|[Co(0.28)|Pd(0.9)]$_N$ |Pd(2) as shown in Fig. 1 inset. The numbers in the parenthesis are in nanometers (nm) and indicate the thickness of each layer (for details see supplementary information). '*N*' is the number of bilayer repeats present in the stack. We have deliberately chosen four different samples, with *N* = 4, 8, 20 and 50 for our study, covering a broad total magnetic thickness range from about 5 nm to about 60 nm, i.e. from well below to well above the penetration depth of the pump laser beam. The pump and probe beams have pulse widths of about 60 fs and about 40 fs, respectively.

Prior to the time-resolved magneto-optical measurement of ultrafast dynamics, we investigated the static properties of the samples using static magneto-optical Kerr effect (S-MOKE) at room temperature. The applied magnetic field was applied in the direction perpendicular to the film plane which helped to measure Kerr rotation in polar geometry. The primary focus of this article is to investigate the time-resolved ultrafast demagnetization of the multi-layered structures with high perpendicular anisotropy. We used a time-resolved magneto-optical Kerr effect (TR-MOKE) magnetometer (see supplementary information) to measure the Kerr rotation, which is proportional to the change in magnetization (*M*), as a function of time. A variable external magnetic field is applied in the out-of-plane direction. Thus, all the measurements, apart from examining the effect of domain structures, are performed in remanent state of the samples which avoids any difference due to the domain structures. Intending to investigate the ultrafast demagnetization, the Kerr rotation trace has been detected over a broader time scale, thus capturing both the ultrafast demagnetization within a few hundreds of femtoseconds as well as the subsequent relaxation within a few picoseconds.

## III. RESULTS AND DISCUSSIONS

Figure 1 shows the hysteresis loops measured in polar Kerr geometry for different samples. It clearly shows that the squareness of the loop decreases and the magnetic saturation field increases, indicating a change of the ground state domain structure towards labyrinth/stripe domains as a function of number of bilayer repeats. This effect is well known and has been studied extensively earlier for similar systems.[30] The raw experimental data obtained from TRMOKE are fitted with a phenomenological expression, obtained by solving the equations from the three-temperature model[31,32], to extract the ultrafast demagnetization and fast relaxation times (see supplementary information). Figure 2 shows the typical ultrafast demagnetization curves measured for different samples with $N$ = 4, 50 (for $N$ = 8, 20, see supplementary information). For each of them, the pump fluence has been varied over a large range, upto a value lower than damage threshold. Although the initial demagnetization part is similar for all the samples over a large range of pump fluence, the follow-up or recovery part is significantly different for the samples with higher $N$ at higher pump fluences. We have fitted the obtained experimental data with the well-known three temperature model expression as given in Eq. 1 below

$$-\Delta\theta_k = \{[\frac{A_1}{(t/t_0+1)^{0.5}} - \frac{(A_2\tau_E - A_1\tau_M)}{\tau_E - \tau_M}e^{-t/\tau_M} - \frac{\tau_E(A_1 - A_2)}{\tau_E - \tau_M}e^{-t/\tau_E}]H(t) + A_3\delta(t)\} \otimes G(t) \quad (1)$$

obtained by solving the energy rate equation in between three different degrees of freedom, e.g. electron, spin and lattice under low pump fluence condition. Although the formula is derived under low fluence condition, it is valid for fluence values similar to ours for extraction of the demagnetization time[31]. $A_1$, $A_2$ and $A_3$ are constants related to different amplitude of the magnetization. $H(t)$, $G(t)$ and $\delta(t)$ are the Heaviside step function, Gaussian laser pulse and Dirac delta function, respectively. $\tau_M$ and $\tau_E$ are the demagnetization time and fast relaxation time, respectively. The convolution of the exponential decay function with the Gaussian laser pulse with 120 fs of full width at half maxima helps in determining an accurate value of the demagnetization time. The experimental data are fitted with the above equation and both demagnetization times, relaxation times are extracted.

The phenomenological fitting shows a slight change in the demagnetization time for $N = 20$ as compared to $N = 4, 8$. The change in demagnetization time with fluence for the samples with $N = 4, 8$ and $20$ is small (~35 - 40 fs). In some previous reports, a similar but greater increment in demagnetization time was found in case of 3d transition metals. The increasing pump fluence gradually pushes the electron temperature closer to the Curie temperature. This leads to enhanced critical magnetic fluctuations and gradual slowing down of the demagnetization process[33,34]. The change in demagnetization time with fluence becomes more significant and prominent for $N = 50$. Careful observation of the demagnetization traces for $N = 50$ clearly shows that it constitutes of two different steps, which is absent for both the lower thicknesses ($N = 4$ and $N = 8$) and the lower fluence conditions (32, 45 mJ/cm$^2$). The additional step leads to the huge enhancement in the demagnetization time for large thickness ($N = 50$) and large fluences ($\geq 50$ mJ/cm$^2$).

Although the pump fluences are varied in a similar fashion for all the samples, the resulting demagnetization curves exhibit significantly different trend. Earlier theoretical investigation at very high pump fluence revealed similar slower recovery process with much less pronounced dip compared to the final demagnetized state[35]. For the lower thicknesses of the samples, the nature of demagnetization as well as the demagnetization time remains unchanged with fluence. However it changes drastically with an additional step of demagnetization for the higher thickness sample with $N=50$. Kuiper *et al.* theoretically shows that for higher thickness of the sample, the demagnetization process is significantly different than of thinner sample[35]. For lower fluences, the samples show the typical SFS induced ultrafast demagnetization timescale for 3d ferromagnetic materials as expected. But, for higher applied pump fluence, the observation of an additional slower demagnetization step indicates the possible occurrence of a type-II demagnetization process where the recovery is much slower, similar to some of the previous studies[15,18,34]. Due to a weak coupling between the electron and spin system in a material, the energy transfer rate becomes slower. As a result, the spin system cannot follow the sudden rise in electronic temperature, and does not attain the equilibrium in hundreds of femtoseconds time scale. The resultant demagnetization becomes slower for these specific materials with weak electron-spin coupling. The samples having same elemental composition should have equal coupling strength in between electron and spin systems and hence, all of them should exhibit type-II or slower observed here, is found to be strongly dependent on the thickness as well as the pump fluence. Hence, it rules out the possibility of weak electron-spin coupling as a reason behind the slowing down of demagnetization in our case. Some recent studies explored the possibility of generation of interlayer spin current and heat current transfer in these kinds of layered structures and its effect on the dynamics[11,13-14,18,23-24]. Hence, to develop a deeper understanding on the slower demagnetization part, we studied the effect of ultrafast demagnetization dynamics as a function of number of bilayers in the sample stack.

Figure 3(a) presents the ultrafast demagnetization traces for all the four samples for the highest applied pump fluence. Using three temperature modelling, we analyse and fit all the traces to extract the demagnetization times. In Fig. 3(b), the variation in demagnetization time with fluence distinctly shows that the nature of the demagnetization changes (demagnetization time changes from femtoseconds to sub picoseconds) as we increase the effective thickness of the sample stack (i.e. for higher number of bilayers). Surprisingly we do not observe any trace of a second step of demagnetization for lower thickness samples in an exactly same experimental arrangement. This observation triggers the idea of the generation of a passive flow of excitation in this multilayer stack. A heat current can indirectly trigger ultrafast demagnetization without any direct interaction in between the laser pulse and the ferromagnetic material. For the samples with smaller number of bilayers $N$ (i.e. smaller effective thickness), both the incident pump and probe pulse penetrate down to the bottom of the stack. It leads to a direct interaction between the spin system and the pump pulse across the whole thickness of the sample, which results in nearly uniform ultrafast demagnetization. The response from the whole sample is consecutively detected by the probe pulse as shown in Fig. 3(c). In this case, only direct process (e.g. SFS) contributes to the ultrafast demagnetization process. In the second case the value of N for the sample is chosen to be much higher, so that the effective sample thickness becomes much larger than the optical penetration depth of the 400 nm pump pulse (Fig. 3(d)). Thus, the pump beam can directly interact only with the spins in the top few layers. It leads to the demagnetization of those layers and generates a sudden non-equilibrium diffusive heat current. It diffuses along the thickness, flowing from the top towards the bottom of the sample. This flow of heat carries enough energy to excite the spin system passively in those ferromagnetic layers, where there is no direct laser excitation because of the limited penetration depth of the pump beam. This causes further demagnetization on a longer time scale due to indirect excitation, which is then governed by the diffusive regime and is detected by the probe beam (800 nm), which has a significantly higher penetration depth than the pump beam (400 nm). Recently, a report by Vodungbo et al.[37] shows that similar indirect excitation can lead to efficient ultrafast demagnetization. However, in that case the excitation mechanism is an altogether different scenario. Here, we have demonstrated a co-excitation of direct and indirect ultrafast demagnetization rather than only indirect demagnetization. Moreover, we have demonstrated a novel way to control the contribution of indirect excitation by using the pump fluence. According to the heat current mechanism, one may also observe indirect excitation in a single thick ferromagnetic layer under favourable conditions.

Earlier, to isolate the effect of a passive (i.e. heat current) and indirect interaction from other direct one, researchers studied various samples systems using several complex experimental geometries. However, so far, the experiments involve either very complicated sample stacks or experimental conditions, which make the detection and isolation of heat currents very difficult. Here, we have detected and confirmed the existence of a heat current induced indirect excitation using a simple experimental scheme of pump wavelength variation which is discussed later in this manuscript. In our case, the direct access to the detection of the indirect heat current flow lies in the different penetration depth of the pump versus probe laser beam inside the sample. It is worth mentioning that our specific experimental design (larger pump spot size than probe) diminishes the effect of heat flow in the sample plane. As the multilayer has anisotropic thermal conductivity (greater in lateral than in normal to the plane), it seems that lateral heat flow will be significant. However, even if the lateral heat flow is ten times faster than the normal one, the contribution from in plane flow will be negligible because thickness of the samples is 1000 times smaller than both the spot sizes. To eliminate the role of inter domain spin transport[37-39], we verified ultrafast demagnetization at several magnetic field values (see supplementary information). Next, we explain the reason for observing two-step demagnetization only at higher fluence. During the diffusion towards the bottom of the sample, the heat current intensity decreases as a result of scattering and absorption. Hence, only a fraction of the initially generated heat current survives for the passive excitation of ultrafast demagnetization at the bottom. On the other hand, the initial intensity of the heat generated strongly depends on the number of interacting pump photons, which in turn is proportional to the pump fluence. Therefore, an increasing fluence effectively enhances the intensity of the initial heat current generation and thereby transferring more heat current for passive or indirect excitation of ultrafast demagnetization.

In order to confirm our speculation, we measured the response of the ultrafast magnetization quenching of the samples for different pump wavelength. Figure 4 shows the Kerr rotation traces corresponding to the magnetization variation for different pump wavelength at a fixed applied fluence for the sample with $N = 50$. The pump pulse width remains nearly constant (about 60 fs) over the whole wavelength range. The variation of the pump wavelength changes the penetration depth, i.e. the extent of direct interaction. The penetration depth corresponding to the wavelengths 400, 480, 550, 600, 650 and 690 nm are estimated[40,41,42] to be 18.0, 21.0, 23.0, 24.0, 24.6, 25.5, 26.2 nm, respectively and that of the probe wavelength (800 nm) is 28.0 nm. Hence, depending on the pump wavelength the contribution to the demagnetization due to indirect excitation should change. Here, we have increased the pump wavelength and show that the demagnetization, purely due to the indirect excitation systematically decreases. This is clearly imprinted in the change in demagnetization time with excitation wavelength. The pulse width of the laser beam of different wavelength remains nearly constant over the range (~400-800 nm). Hence, the effect of wavelength variation on the demagnetization time can be ruled out. Actually, the increasing pump wavelength increases the penetration depth and thus reduces the indirect excitation volume as well. This results in decrement of the demagnetization time with increment in wavelength as shown in Fig. 4. Inset schematic illustrates the aforementioned phenomena. It depicts the penetration depth as well as the direct interaction volume for both pump and probe beam inside the sample. As a result of increasing pump wavelength, (i.e. from 400 nm to 690 nm), the penetration depth increases, which in turn, reduces the strength of indirect excitation. However, the probe detection volume remains the same in both cases. As a result, the two-step demagnetization turns into a single step demagnetization, which can be clearly seen in Fig. 4.

Magnetic multilayers having a strong perpendicular magnetic anisotropy and significantly large thickness usually exhibit a labyrinth stripe domain structure with a domain width of around one hundred to a few hundred nanometres. As a result of the direct transfer of spin angular momentum in between these alternate nanometric magnetic domains with opposite magnetization state, the ultrafast demagnetization time can be significantly modified. To examine the impact of domain formation, we measured the ultrafast demagnetization of the 50-repeat sample ($N = 50$) at several applied magnetic fields during the domain reversal process, which is shown in Fig. 5. It is interesting to note that we did not observe any difference in between the demagnetization traces as a function of applied magnetic field. Even the absence (saturated state = remanent state) and presence (unsaturated) of stripe domain did not affect the demagnetization time characteristics. It is worth mentioning the fact that both domain width (> 100 nm) and domain wall width are significantly larger than the spin diffusion length in Pd. That eliminates any role of domain here. This clearly rules out the possible role of inter-domain spin transport in our case.

## IV. CONCLUSION

In summary, we have investigated the ultrafast demagnetization dynamics in a magnetic multilayer with high PMA and unveil a new way of understanding the basic underlying mechanisms. We found a sudden rise in the ultrafast demagnetization time with a transition from single-step demagnetization to a two-step demagnetization process for higher applied pump fluence and thicker samples. A systematic in-depth investigation reveals that the process of ultrafast magnetization quenching is, also governed by an indirect excitation via diffusive heat current transport in addition to direct excitation. Furthermore, we measured the ultrafast demagnetization by systematically changing the sample thickness and the excitation wavelength. Although the direct observation and isolation of the contribution due to a diffusive heat energy transport from other direct contributions has been quite complicated and challenging, we present here a clear and simple pathway to study the transition from a direct excitation to an indirect excitation dominated regime and identification of the heat current contribution. Here, we exploit a simple concept of direct scaling of skin depth with excitation wavelength. Our study enlightens a new and simple method to understand the long debated ultrafast demagnetization mechanism and confirms the possibility of pure indirect excitation. This is an important step towards a deeper understanding of the mechanisms in such complex multilayer systems and towards putting such systems forward for device application. We hope further extended and systematic studies on a series of single and multilayer ferromagnetic thin films will firmly establish this phenomenon.

## ACKNOWLEDGMENTS

We acknowledge the financial assistance from Department of Science and Technology, Govt. of India under grant No. SR/NM/NS-09/2011 and S. N. Bose National Centre for Basic Sciences under project no. SNB/AB/12-13/96 and SNB/AB/18-19/211. S.P. acknowledges DST, Govt. of India for INSPIRE fellowship (IF04999).*abarman@bose.res.in

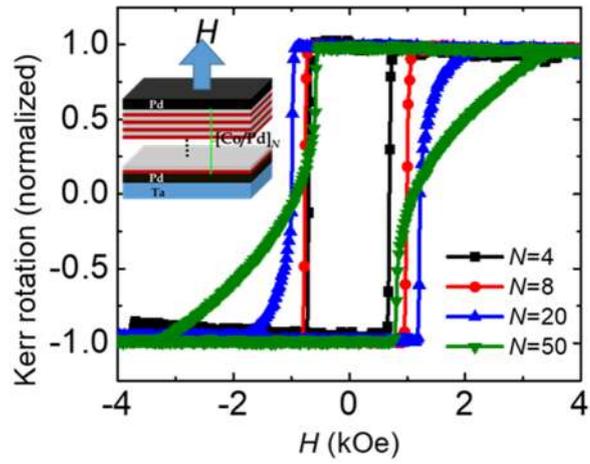

FIG. 1. Hysteresis loops measured in polar Kerr geometry for all four samples with number of bilayers N = 4,8,20, 50. Inset shows the applied magnetic field direction with respect to sample stack.

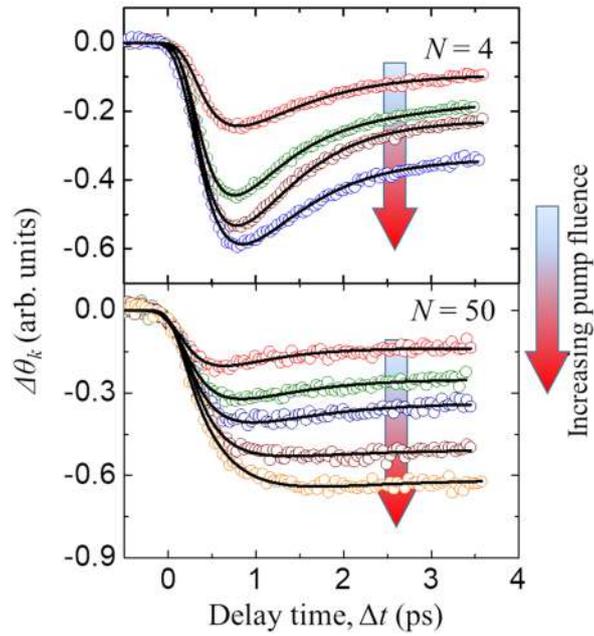

FIG. 2. Change in Kerr rotation (i.e. ultrafast demagnetization) traces for samples with $N=4$ and 50 at several applied pump fluences of 19, 32, 45, 57 mJ/cm$^2$, respectively. An additional set of data for $N = 50$ is measured at a fluence of 70 mJ/cm$^2$. Pump and probe beam are of 400 and 800 nm, respectively.

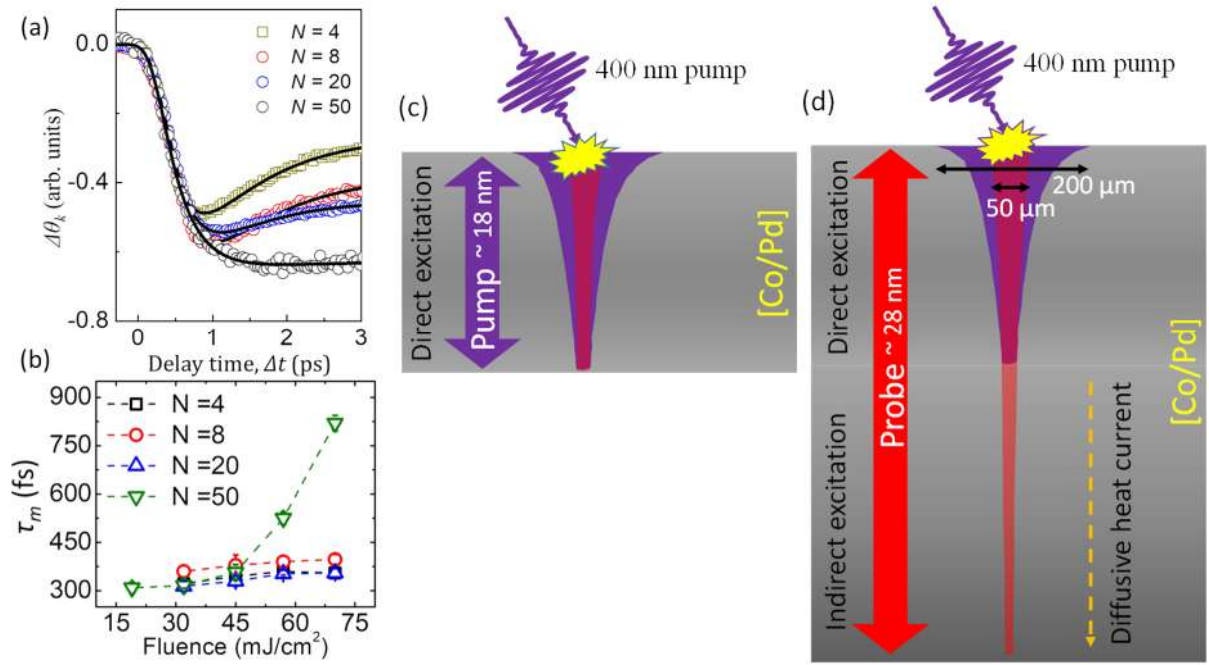

FIG. 3. (a) Kerr rotation traces at a fixed pump (70 mJ/cm$^2$) and probe fluence, (b) demagnetization time ($\tau_m$) versus pump fluence, for all the samples. Penetration of both pump (blue) and probe (red) laser in the sample; (c) only direct excitation in thinner sample where pump beam (~400 nm) and probe beam (~800 nm) both reaches to the bottom of the sample, (d) both direct and indirect excitation in thicker sample where pump (~400 nm) does not reach to bottom most part of the sample but probe beam (~800 nm) reaches.

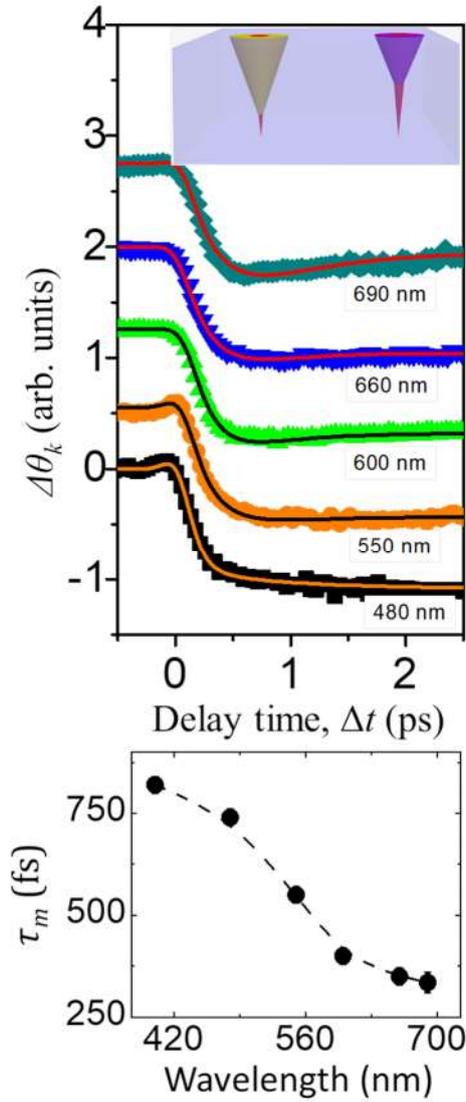

FIG. 4. Kerr rotation traces for different pump wavelength (values written in the figure) excitation with a fixed probe wavelength at 800 nm and at a fixed pump fluence of 70 mJ/cm$^2$. Inset shows the change in excitation volume of pump beam due to variation in wavelength (top). Plot of extracted demagnetization time ($\tau_m$) versus wavelength showing a gradually diminishing effect of indirect excitation. The dashed line in the plot is just a "guide to eye line" (bottom).

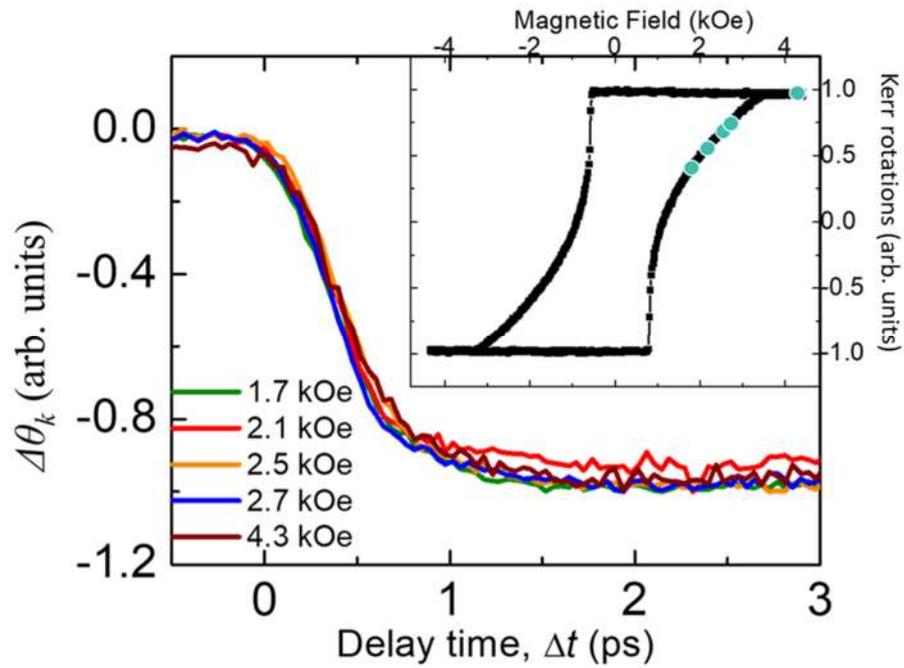

FIG. 5. Kerr rotation traces for the sample with $N = 50$ at various applied magnetic field for pump fluence of 70 mJ/cm$^2$ (values of external applied magnetic field are depicted on the inset hysteresis loop). Pump and probe wavelength was fixed at 400 nm and 800 nm, respectively.